\documentclass[preprint]{aastex}

\usepackage{epsf}

\newcommand{\bl}{\mbox{\boldmath$l$}}

\newcommand{\bG}{\mbox{\boldmath$G$}}
\newcommand{\bT}{\mbox{\boldmath$T$}}

\begin{document}

\title{Secular interactions between inclined planets and a gaseous disk}
\shorttitle{Secular planet--disk interactions}
\author{S. H. Lubow\altaffilmark{1,2}
 and G. I. Ogilvie\altaffilmark{1,2}}
\altaffiltext{1}{Space Telescope Science Institute,
  3700 San Martin Drive, Baltimore, MD 21218}
\altaffiltext{2}{Institute of Astronomy, University of Cambridge,
  Madingley Road, Cambridge CB3 0HA, UK}
\shortauthors{Lubow \& Ogilvie}

\begin{abstract}

  In a planetary system, a secular particle resonance occurs at a
  location where the precession rate of a test particle (e.g. an
  asteroid) matches the frequency of one of the precessional modes of
  the planetary system.  We investigate the secular interactions of a
  system of mutually inclined planets with a gaseous protostellar disk
  that may contain a secular nodal particle resonance.  We determine
  the normal modes of some mutually inclined planet-disk systems. The
  planets and disk interact gravitationally, and the disk is
  internally subject to the effects of gas pressure, self-gravity, and
  turbulent viscosity.  The behavior of the disk at a secular
  resonance is radically different from that of a particle, owing
  mainly to the effects of gas pressure.  The resonance is typically
  broadened by gas pressure to the extent that global effects,
  including large-scale warps, dominate. The standard resonant torque
  formula is invalid in this regime.  Secular interactions cause a
  decay of the inclination at a rate that depends on the disk
  properties, including its mass, turbulent viscosity, and sound
  speed.  For a Jupiter-mass planet
  embedded within a minimum-mass solar nebula having typical parameters,
  dissipation within the disk is sufficient to stabilize the system
  against tilt growth caused by mean-motion resonances.

\end{abstract}

\keywords{accretion, accretion disks --- celestial mechanics ---
  hydrodynamics --- planets and satellites: general --- solar system:
  general --- waves}

\section{Introduction}

The interaction between a young planetary system and its
protoplanetary disk likely plays an important role in determining the
orbital properties of the planets.  Such interactions may be important
for understanding the observed orbital properties of extra-solar
planets (e.g. Marcy et al. 1999).  Resonances within a gaseous disk
likely play a key role in determining planetary eccentricities and
inclinations. Much of the previous work, starting with Goldreich \&
Tremaine (1980), has emphasized the effects of mean-motion disk
resonances, which involve frequencies that are comparable to the
orbital frequencies of the planets.

Another potentially important class of resonances occurs where there
is a matching of the precession frequency of a test particle (e.g. an
asteroid) with the frequency of one of the precessional modes of the
planetary system. The frequencies involved are much lower than in the
mean-motion case.  At such a resonance, the motion of a test particle
can be strongly driven by the planets, resulting in a high orbital
inclination (for a nodal resonance) or eccentricity (for an apsidal
resonance).  For example, in the solar system there is an important
secular resonance that occurs near 2 AU due to driving involving
Jupiter and Saturn.  This resonance is believed to be associated with
the inner truncation of the asteroid belt (Tisserand 1882).

It was recognized by Ward, Colombo, \& Franklin (1976) that these
resonances must have swept across portions of the early solar system
owing to the effects of the gaseous disk, even if the planets do not
migrate.  The reason is that even a minimum-mass solar nebula can have
an important influence on the relevant precession rates.  As the
nebula disperses, the precession rates vary, along with the resonance
locations (Ward 1981). Regions through which the resonances sweep may
be driven into significantly eccentric or inclined orbits, such as are
observed among the asteroids today (e.g. Nagasawa, Tanaka, \& Ida
2000).  The resonances can also have an important bearing on the
conditions for planet formation by stirring the planetesimal disk.

However, collective effects can be important within the disk and may
even result in the development of density waves or bending waves.
Consequently, the response is less localized than would be indicated
by test particles, as was recognized by Ward \& Hahn (1998) and
Tremaine (1998).  These investigations concentrated on the collective
effects of self-gravity in particle disks.

Secular interactions involving the gas disk of a young planetary
system are potentially of importance, since much more mass is
contained in the gas disk than in the planetesimal disk.  The disk is
a fluid body and is capable of wave motions through (at least)
compressive, inertial, buoyancy, and self-gravitational forces.  It is
also capable of dissipating energy through a turbulent effective
viscosity or through shocks.  It is therefore important to understand
the dynamical response of the disk to secular forcing and the
implications for the evolution of the planetary orbits.

The purpose of this paper is to investigate the outcome of secular
interactions in mutually inclined planet-disk systems.  The relevant
waves in the disk are long-wavelength bending waves with azimuthal
wavenumber $m=1$.  The equations governing such waves in a
protostellar disk are fairly well established, while those for
long-wavelength eccentric density waves have received less attention
(although see Ogilvie 2001).  The importance of understanding the
global $m=1$ response of the disk to secular perturbations has been
emphasized by Tremaine (1998).

The general outline of the paper is as follows.  Sections 2--5
describe the physical model for nodal secular interactions and
formulate the normal mode analysis.  Sections 6 and 7 explore the
limit of sufficiently slow modes, when the disk responds nearly
rigidly.  Section 8 describes how the effects of mean-motion
resonances are included in the normal mode analysis.  Sections 9 and
10 describe numerical results for Jupiter interacting with the solar
nebula.  Section 11 provides a simple scaling analysis based on a
nearly rigid tilt model.  Section 12 discusses the relationship of the
current approach with that of earlier work involving WKB theory.
Section 13 describes the analysis of a close-orbiting planet in the
central hole of a disk.  Section 14 discusses Jupiter and Saturn
interacting with the solar nebula at the $\nu_{16}$ secular resonance.

\section{Modeling secular interactions}

\label{Modeling secular interactions}

Planets experience secular gravitational interactions on time-scales
much longer than their orbital periods.  On such time-scales the
planets may be considered as continuous rings representing their
average mass density.  These rings are in general elliptical and
mutually inclined.  Gravitational interactions between the rings lead
to apsidal and nodal precession of the orbits.

In the case of small eccentricities and inclinations, the secular
evolution of a system of $n$ planets can be described in terms of
normal modes of a linear dynamical system, $n$ modes for eccentricity
and $n$ for inclination (e.g. Murray \& Dermott 1999).  The
eigenvector of a mode describes the relative distribution of
eccentricity or inclination among the planetary orbits, while the
eigenfrequency is the rate at which the pattern precesses.

In typical situations involving mean-motion resonances (Lindblad or
vertical resonances), if a particle resonance is located at a certain
radius in a fluid disk, the disk responds by launching a wave that
carries energy and angular momentum away from the resonance.  Such
waves are usually assumed implicitly to be damped through some
dissipative process before reaching an edge of the disk from which
they might otherwise reflect.  The gravitational torque between the
disk and the perturber can then be calculated from a standard formula
evaluated at the location of the resonance (Goldreich \& Tremaine
1979).

However, in the case of a secular resonance, the frequency may be so
small that the wavelength would be comparable to (or may even exceed)
the size of the disk.  In that case the global response of the disk
must be computed, including explicit dissipation and the correct
boundary conditions.  The exact location of the resonance then ceases
to have great importance and the new possibility of a global secular
resonance arises.

We focus on the case of inclination, so that the relevant waves in the
disk are long-wavelength bending waves with azimuthal wavenumber
$m=1$.  As noted above, the equations governing such waves in a
protostellar disk are fairly well established.

\section{Basic equations}

\label{Basic equations}

Let $(r,\phi,z)$ be cylindrical polar coordinates with origin at the
central object, of mass $M_*$.  We consider a protostellar disk of
semi-thickness $H(r)$, for which the turbulent viscosity parameter
$\alpha$ satisfies the condition $\alpha \la H/r$.  The linearized
equations for bending waves in such a disk have been derived in
several papers (Papaloizou \& Lin 1995; Masset \& Tagger 1996;
Demianski \& Ivanov 1997).  We present them in the form (Lubow \&
Ogilvie 2000)
\begin{equation}
  \label{W1}
  \Sigma r^2\Omega{{\partial W}\over{\partial t}}=
  {{1}\over{r}}{{\partial{\cal G}}\over{\partial r}}+T,
\end{equation}
\begin{equation}
  \label{G1}
  {{\partial{\cal G}}\over{\partial t}}+
  \left({{\kappa^2-\Omega^2}\over{\Omega^2}}\right)
  {{i\Omega}\over{2}}{\cal G}+\alpha\Omega{\cal G}=
  {{{\cal I}r^3\Omega^3}\over{4}}{{\partial W}\over{\partial r}}.
\end{equation}
Here $W(r,t)=l_x+il_y$ is the complex tilt variable, which describes
the warped shape of the disk.  Essentially, the disk can be thought of
as a continuum of concentric circular rings with radii $r$ and unit
normal vectors $\bl(r,t)$.  Also ${\cal G}(r,t)=G_x+iG_y$ is the
complex internal torque variable.  In a warped disk there is a
horizontal internal torque $2\pi\bG(r,t)$ that is mediated by
horizontal motions that are proportional to the distance $z$ from the
midplane.  These stresses are responsible for the propagation of
bending waves.  Similarly $T(r,t)=T_x+iT_y$ is the complex horizontal
external torque density acting on the disk.  Finally $\Sigma(r)$ is
the surface density, $\Omega(r)$ the orbital angular velocity,
$\kappa(r)$ the epicyclic frequency and ${\cal I}(r)$ the second
vertical moment of the density.  These are defined by
\begin{equation}
\label{calI}
  \Sigma=\int\rho\,dz,\qquad
  {\cal I}=\int\rho z^2\,dz,
\end{equation}
\begin{equation}
  r\Omega^2={{\partial\Phi}\over{\partial r}}\Bigg|_{z=0},\qquad
  \kappa^2=4\Omega^2+2r\Omega{{d\Omega}\over{dr}},
\end{equation}
where $\Phi(r,z)$ is the (axisymmetric component of the) gravitational
potential experienced by the disk.

Consider a thin, uniform circular ring of mass $m_i\ll M_*$ and radius
$r_i$, representing either an annulus of the disk or the time-average
of a planetary orbit.  Its contribution to the potential in the plane
$z=0$ is
\begin{equation}
  \Phi_i=-{{Gm_i}\over{2\pi}}\int_0^{2\pi}
  (r^2+r_i^2-2rr_i\cos\phi)^{-1/2}\,d\phi.
\end{equation}
Therefore the angular velocity and epicyclic frequency experienced by
the ring are obtained from summations over all other such rings, in
the forms
\begin{equation}
  \Omega^2_i={{GM_*}\over{r_i^3}}+\sum_{j\ne i}{{2Gm_j}\over{r_ir_j}}
  \left[K_0(r_i,r_j)-{{r_j}\over{r_i}}K_1(r_i,r_j)\right],
\label{Omega2}
\end{equation}
\begin{equation}
  \kappa^2_i={{GM_*}\over{r_i^3}}+\sum_{j\ne i}{{2Gm_j}\over{r_ir_j}}
  \left[K_0(r_i,r_j)-{{2r_j}\over{r_i}}K_1(r_i,r_j)\right],
\label{kappa2}
\end{equation}
where $K_0$ and $K_1$ are symmetric kernels with dimensions of inverse
length, given by
\begin{equation}
\label{km}
  K_m(r_i,r_j)={{r_ir_j}\over{4\pi}}\int_0^{2\pi}
  (r_i^2+r_j^2-2r_ir_j\cos\phi)^{-3/2}\cos^m\phi\,d\phi.
\end{equation}
Hereafter we write $K_1$ simply as $K$.

Now consider the tilt interaction between two such rings of masses
$m_i$ and $m_j$, radii $r_i$ and $r_j$, and tilt vectors $\bl_i$ and
$\bl_j$.  For small relative inclinations, the gravitational torque
exerted by ring $j$ on ring $i$ is
\begin{equation}
  \bT_{ji}=Gm_im_jK(r_i,r_j)\,\bl_i\times\bl_j.
\end{equation}
This expression can be obtained by averaging the torque exerted by a
point mass on a circular ring (Lubow \& Ogilvie 2000) over the orbital
motion of the point mass.  Similar expressions can be found in studies
of galactic warps (e.g. Sparke \& Casertano 1988).  In the complex
notation, for small inclinations from the $xy$-plane, the total
external torque acting on ring $i$ is
\begin{equation}
  T_i=\sum_{j\ne i}Gm_im_jK(r_i,r_j)i(W_j-W_i).
\end{equation}

The kernels can also be written as
\begin{equation}
  K_m(r_i,r_j)={{r_<}\over{4r_>^2}}
  b_{3/2}^{(m)}\left({{r_<}\over{r_>}}\right),
\end{equation}
where $r_>=\max(r_i,r_j)$, $r_<=\min(r_i,r_j)$ and $b_\gamma^{(m)}$ is
the Laplace coefficient.  In practice we evaluate these in terms of
elliptic integrals using Carlson's algorithms (Press et al. 1992).

These kernels diverge as $|r_i-r_j|\to0$.  In reality the
gravitational interaction remains bounded because of the non-zero
vertical thickness of the rings, which we have neglected.  To take
account of this in an approximate way, we soften the kernels by
replacing
\begin{equation}
\label{ksoft}
  K_m(r_i,r_j)\mapsto K_m\left({{r_i+r_j-h}\over{2}},
  {{r_i+r_j+h}\over{2}}\right),
\end{equation}
whenever $|r_i-r_j|<h$.  Here the smoothing length $h$ is to be
understood as an approximate measure of the thickness of the rings.

\section{Coupled systems of planets and disks}

\label{planets and disks}

We consider a general system consisting of multiple planets embedded
in a gaseous disk. Provided the planets create gaps in the disk, the
disk then becomes partitioned into a set of disks. The disks and
planets interact through gravity.

We consider planets $i=1,2,\dots,n_{\rm p}$ of masses $m_{{\rm p}i}$
in circular orbits of radii $r_{{\rm p}i}$ and angular velocities
$\Omega_{{\rm p}i}$.  We also consider disks $k=1,2,\dots,n_{\rm d}$
of inner radii $a_k$ and outer radii $b_k$.  The inclinations of the
planetary orbits are described by the tilt variables $W_{{\rm
    p}i}(t)$. The tilt within a disk is described by $W_{{\rm d}k}(r,
t)$.  We assume throughout that the inclinations are small, so that
linear theory applies.

If only secular nodal interactions are considered, the planets can be
treated as inclined circular rings interacting with each other and
with the disks through the torques described in Section~\ref{Basic
  equations}.  We return later (Section~\ref{Mean motion}) to the
effect of mean-motion resonances on tilt evolution.  However, we
assume throughout that any evolution of the surface density of the
disk, or of the semi-major axes of the planetary orbits, may be
neglected.

The secular equations for the planets are then of the form
\begin{eqnarray}
  m_{{\rm p}i}r_{{\rm p}i}^2\Omega_{{\rm p}i}{{dW_{{\rm p}i}}\over{dt}}&=&
  \sum_{j\neq i}Gm_{{\rm p}i}m_{{\rm p}j}K(r_{{\rm p}i},r_{{\rm p}j})
  i(W_{{\rm p}j}-W_{{\rm p}i})\nonumber\\
  &&+\sum_k\int_{a_k}^{b_k}Gm_{{\rm p}i}\Sigma_k K(r,r_{{\rm p}i})
  i(W_{{\rm d}k}-W_{{\rm p}i})\,2\pi r\,dr,
  \label{planet}
\end{eqnarray}
while the equations for the disks are
\begin{eqnarray}
  \Sigma_k r^2\Omega{{\partial W_{{\rm d}k}}\over{\partial t}} &=&
  {{1}\over{r}}{{\partial{\cal G}_k}\over{\partial r}}+
  \sum_iGm_{{\rm p}i}\Sigma_k K(r,r_{{\rm p}i})i
    (W_{{\rm p}i}-W_{{\rm d} k})\nonumber\\
  &&+\sum_\ell\int_{a_\ell}^{b_\ell}
     G\Sigma_k\Sigma'_{\ell} K(r,r')i(W'_{{\rm d} {\ell}}-W_{{\rm d}k})
     \,2\pi r'\,dr',
  \label{disk}
\end{eqnarray}
\begin{equation}
  {{\partial{\cal G}_k}\over{\partial t}}+
  \left({{\kappa^2-\Omega^2}\over{\Omega^2}}\right)
  {{i\Omega}\over{2}}{\cal G}_k+\alpha\Omega{\cal G}_k=
  {{{\cal I}_kr^3\Omega^3}\over{4}}
  {{\partial W_{{\rm d} k}}\over{\partial r}},
  \label{torque}
\end{equation}
where $\Sigma'_\ell=\Sigma_\ell(r')$ and $W'_{{\rm d}\ell}=W_{{\rm
    d}\ell}(r',t)$.  The final term in equation (\ref{disk})
represents the self-gravitation of each disk and the gravitational
interactions between disks.

The (complex) total horizontal angular momentum of the system,
\begin{equation}
  \sum_im_{{\rm p}i}r_{{\rm p}i}^2\Omega_{{\rm p}i}W_{{\rm p}i}+
  \sum_k\int_{a_k}^{b_k}\Sigma_k r^2\Omega W_{{\rm d}k}\,2\pi r\,dr,
\end{equation}
is exactly conserved if, as we assume, the boundary conditions
\begin{equation}
  {\cal G}_k(a_k,t)={\cal G}_k(b_k,t)=0
\end{equation}
hold at all times $t$.  There is also a conservation law for the
  vertical angular momentum, which we derive in Appendix~A.

\section{Normal modes}

\label{Normal modes}

We have a set of coupled integro-differential equations that are
linear and homogeneous.  Solutions may be sought in the form of normal
modes,
\begin{equation}
  W_{{\rm p}i}(t)=\tilde W_{{\rm p}i}\,e^{i\omega t},\qquad
  W_{{\rm d}k}(r,t)=\tilde W_{{\rm d}k}(r)\,e^{i\omega t},\qquad
  {\cal G}_k(r,t)=\tilde{\cal G}_k(r)\,e^{i\omega t},
\end{equation}
where $\omega$ is a complex frequency eigenvalue.  The real part of
$\omega$ is the precession rate of the mode, while the imaginary part
is the decay rate.  Substituting this and omitting the tildes, we
obtain the integral and integro-differential equations
\begin{eqnarray}
  i\omega m_{{\rm p}i}r_{{\rm p}i}^2\Omega_{{\rm p}i}W_{{\rm p}i}&=&
  \sum_{j\neq i}Gm_{{\rm p}i}m_{{\rm p}j}K(r_{{\rm p}i},r_{{\rm p}j})
  i(W_{{\rm p}j}-W_{{\rm p}i})\nonumber\\
  &&+\sum_k\int_{a_k}^{b_k}Gm_{{\rm p}i}\Sigma_k K(r,r_{{\rm p}i})
  i(W_{{\rm d}k}-W_{{\rm p}i})\,2\pi r\,dr, \label{wpnm}\\
  i\omega\Sigma_k r^2\Omega W_{{\rm d} k}&=&
  {{1}\over{r}}{{d{\cal G}_k}\over{dr}}+
  \sum_iGm_{{\rm p}i}\Sigma_k K(r,r_{{\rm p}i})i
    (W_{{\rm p}i}-W_{{\rm d} k})\nonumber\\
  &&+\sum_\ell\int_{a_\ell}^{b_\ell}
     G\Sigma_k\Sigma'_\ell K(r,r')i(W'_{{\rm d}\ell}-W_{{\rm d}k})
    \,2\pi r'\,dr',
  \label{iomegaw}
\end{eqnarray}
\begin{equation}
  i\omega{\cal G}_k+   \left({{\kappa^2-\Omega^2}\over{\Omega^2}}\right)
  {{i\Omega}\over{2}}{\cal G}_k+\alpha\Omega{\cal G}_k =
  {{{\cal I}_kr^3\Omega^3}\over{4}}{{dW_{{\rm d}k}}\over{dr}},
  \label{dwdr}
\end{equation}
subject to the boundary conditions
\begin{equation}
  \label{bcs}
  {\cal G}_k(a_k)={\cal G}_k(b_k)=0.
\end{equation}
Frequencies $\Omega$ and $\kappa$ are given by equations (\ref
{Omega2}) and (\ref {kappa2}).  All orbits are assumed to be prograde.
There is always a trivial rigid-tilt mode with $\omega=0$, ${\cal
  G}_k=0$, and $W_{{\rm p}i}=W_{{\rm d} k}={\rm constant}$.

\section{Rigid response}

\label{Rigid response}

Secular modes usually have very long periods.  If a disk is easily
able to maintain radial communication (through pressure, viscosity or
self-gravitation) on this long time-scale, it will participate in such
a mode almost as a rigid body, and $W$ will not vary much across the
disk.

Suppose that each disk indeed behaves rigidly ($W_{{\rm d}k}={\rm
  constant}$).  Then ${\cal G}_k$ can be eliminated by multiplying
equation (\ref{iomegaw}) by $2\pi r$, integrating from $a_k$ to $b_k$,
and using the boundary conditions.  We then find
\begin{equation}
  \omega J_{{\rm p}i}W_{{\rm p}i}=
  \sum_{j\neq i}C^{\rm pp}_{ij}(W_{{\rm p}j}-W_{{\rm p}i})+
  \sum_kC^{\rm pd}_{ik}(W_{{\rm d}k}-W_{{\rm p}i}),
\end{equation}
\begin{equation}
  \omega J_{{\rm d}k}W_{{\rm d}k}=
  \sum_iC^{\rm pd}_{ik}(W_{{\rm p}i}-W_{{\rm d}k})+
  \sum_{\ell\ne k}C^{\rm dd}_{k\ell}(W_{{\rm d}\ell}-W_{{\rm d}k}),
\end{equation}
where
\begin{equation}
  J_{{\rm p}i}=m_{{\rm p}i}r_{{\rm p}i}^2\Omega_{{\rm p}i}
\end{equation}
is the angular momentum of planet $i$, and
\begin{equation}
  J_{{\rm d}k}=\int_{a_k}^{b_k}\Sigma r^2\Omega\,2\pi r\,dr
\end{equation}
is the total angular momentum of disk $k$.  The coupling coefficients
are defined by
\begin{equation}
  C^{\rm pp}_{ij}=Gm_{{\rm p}i}m_{{\rm p}j}K(r_{{\rm p}i},r_{{\rm p}j}),
\end{equation}
\begin{equation}
  C^{\rm pd}_{ik}=\int_{a_k}^{b_k}Gm_{{\rm p}i}\Sigma_k
  K(r,r_{{\rm p}i})\,2\pi r\,dr,
\end{equation}
\begin{equation}
  C^{\rm dd}_{k\ell}=\int_{a_\ell}^{b_\ell}\int_{a_k}^{b_k}
  G\Sigma_k\Sigma'_\ell K(r,r')\,2\pi r\,dr\,2\pi r'\,dr',
\end{equation}
Evidently the disks can be treated formally as additional planets if
they behave as rigid bodies.  By introducing Greek indices that run
from $1$ to $n_{\rm p}+n_{\rm d}$, we obtain the algebraic eigenvalue
problem
\begin{equation}
  \omega J_\beta W_\beta =\sum_{\gamma\neq\beta}
  C_{\beta\gamma}(W_\gamma-W_\beta).
\end{equation}
This is formally identical to the multiple-planet problem described
by, e.g., Murray \& Dermott (1999).  Since $C$ is a
symmetric matrix, we have
\begin{equation}
  \omega\sum_\beta J_\beta|W_\beta|^2=-{{1}\over{2}}\sum_{\beta,\gamma}
  C_{\beta\gamma}|W_\beta-W_\gamma|^2.
  \label{rigid_sum}
\end{equation}
Since, further, the coefficients $C_{\beta\gamma}$ are real and
positive, the eigenvalues are all real and non-positive.  Apart from
the trivial rigid-tilt mode, all modes precess retrogradely without
growth or decay.

Suppose the system contains only two components, e.g. one planet
interacting with a connected disk.  The rigid solution satisfies
\begin{equation}
  \left[\matrix{\omega J_1+C&-C\cr -C&\omega J_2+C}\right]
  \left[\matrix{W_1\cr W_2\cr}\right]=\left[\matrix{0\cr 0}\right],
\end{equation}
where $C=C_{12}$ is the coupling coefficient between the planet and
the disk.  The eigenvalues and eigenvectors are
\begin{equation}
  \omega=0,\qquad
  \left[\matrix{W_1\cr W_2\cr}\right]\propto
  \left[\matrix{1\cr 1\cr}\right],
\end{equation}
a trivial solution corresponding to a rigid tilt, and
\begin{equation}
  \omega=-{{C(J_1+J_2)}\over{J_1J_2}},\qquad
  \left[\matrix{W_1\cr W_2\cr}\right]\propto
  \left[\matrix{J_2\cr -J_1\cr}\right],
  \label{non-trivial}
\end{equation}
corresponding to a retrogradely precessing mode in which the tilt of
each component is inversely proportional to its angular momentum (in
order to conserve the total angular momentum).

In a three-component system, e.g. a planet interacting with interior
and exterior disks, the rigid solutions satisfy
\begin{equation}
  \left[\matrix{\omega J_1+C_{12}+C_{13}&-C_{12}&-C_{13}\cr
  -C_{12}&\omega J_2+C_{12}+C_{23}&-C_{23}\cr
  -C_{13}&-C_{23}&\omega J_3+C_{13}+C_{23}}\right]
  \left[\matrix{W_1\cr W_2\cr W_3\cr}\right]=
  \left[\matrix{0\cr 0\cr 0}\right],
\end{equation}
where subscript 1 refers to the planet, and subscripts 2 and 3 to the
inner and outer disks.

In the general case, the eigenvalues and eigenfunctions are
algebraically complicated.  However, consider the case in which the
planet is much less massive than either disk.  The first non-trivial
mode is the equivalent of equation (\ref{non-trivial}), but involving
the two dominant components:
\begin{equation}
  \omega\approx-{{C_{23}(J_2+J_3)}\over{J_2J_3}},\qquad
  \left[\matrix{W_2\cr W_3\cr}\right]\propto
  \left[\matrix{J_3\cr -J_2\cr}\right].
  \label{mode1}
\end{equation}
The planet's tilt is then driven according to
\begin{equation}
  W_1={{C_{12}W_2+C_{13}W_3}\over{\omega J_1+C_{12}+C_{13}}}.
\end{equation}
Secular resonance occurs here if the frequency is such that the
denominator vanishes.

Secular resonance of this type could also occur, for example, in a
system of two massive planets and one low-mass disk.  In this case the
disk would be responding globally as a rigid body.  This is quite
different from the situation in which a secular particle resonance is
located somewhere within a disk.

The second non-trivial mode has
\begin{equation}
  \omega\approx-{{(C_{12}+C_{13})}\over{J_1}},\qquad
  |W_2|,|W_{3}|\ll|W_1|.
  \label{mode2}
\end{equation}
Here only the planet's orbit is significantly tilted, and precesses at
a rate determined by the two disks.

\section{Nearly rigid response}

\label{Nearly rigid response}

The internal torque ${\cal G}_k$ required for a disk to respond
rigidly can be determined from equation (\ref{iomegaw}).  In general,
equation (\ref{dwdr}) will then indicate that a slight warping is in
fact required.  The rigid solution can be understood as the leading
term in an expansion of the solution in powers of a small parameter.
If $\epsilon=H/r$ is the angular semi-thickness of the disk, and
$c_{\rm s}\sim H\Omega$ the isothermal sound speed, the small
parameter here is a characteristic value of
$(\omega/\epsilon\Omega)^2$, or $(\omega r/c_{\rm s})^2$.  This
parameter is small when the disk can communicate effectively, i.e.
when the time taken for a bending wave to propagate across the disk is
less than the wave period.

Now suppose the disk behaves {\it nearly} rigidly, so that we may pose
an expansion in powers of this parameter, of the form
\begin{eqnarray}
  \omega&=&\omega^{(0)}+\omega^{(1)}+\cdots,\\
  W_{{\rm p}i}&=&W_{{\rm p}i}^{(0)}+W_{{\rm p}i}^{(1)}+\cdots,\\
  W_{{\rm d}k}&=&W_{{\rm d}k}^{(0)}+W_{{\rm d}k}^{(1)}(r)+\cdots,\\
  {\cal G}_k&=&{\cal G}_k^{(0)}(r)+{\cal G}_k^{(1)}(r)+\cdots.
\end{eqnarray}
At leading order we obtain the rigid solution described in
Section~\ref{Rigid response}, denoted by the superscript $(0)$.  At
the next order we find
\begin{eqnarray}
  &&i\omega^{(1)}J_{{\rm p}i}W_{{\rm p}i}^{(0)}+
  i\omega^{(0)}J_{{\rm p}i}W_{{\rm p}i}^{(1)}=
  \sum_{j\neq i}Gm_{{\rm p}i}m_{{\rm p}j}K(r_{{\rm p}i},r_{{\rm p}j})
  i(W_{{\rm p}j}^{(1)}-W_{{\rm p}i}^{(1)})\nonumber\\
  &&\qquad+\sum_k\int_{a_k}^{b_k}Gm_{{\rm p}i}\Sigma_k K(r,r_{{\rm p}i})
  i(W_{{\rm d}k}^{(1)}-W_{{\rm p}i}^{(1)})\,2\pi r\,dr, \\
  &&i\omega^{(1)}\Sigma_k r^2\Omega W_{{\rm d}k}^{(0)}+
  i\omega^{(0)}\Sigma_k r^2\Omega W_{{\rm d}k}^{(1)}=
  {{1}\over{r}}{{d{\cal G}_k^{(1)}}\over{dr}}+
  \sum_iGm_{{\rm p}i}\Sigma_k K(r,r_{{\rm p}i})i
    (W_{{\rm p}i}^{(1)}-W_{{\rm d} k}^{(1)})\nonumber\\
  &&\qquad+\sum_\ell\int_{a_\ell}^{b_\ell}
     G\Sigma_k\Sigma'_\ell K(r,r')
  i(W^{\prime(1)}_{{\rm d}\ell}-W_{{\rm d}k}^{(1)})\,2\pi r'\,dr',
\end{eqnarray}
\begin{equation}
  i\omega^{(0)}{\cal G}_k^{(0)}+
  \left({{\kappa^2-\Omega^2}\over{\Omega^2}}\right)
  {{i\Omega}\over{2}}{\cal G}_k^{(0)}+\alpha\Omega{\cal G}_k^{(0)}=
  {{{\cal I}_kr^3\Omega^3}\over{4}}{{dW_{{\rm d}k}^{(1)}}\over{dr}},
\end{equation}
subject to the boundary conditions
\begin{equation}
  {\cal G}_k^{(1)}(a_k)={\cal G}_k^{(1)}(b_k)=0.
\end{equation}
After some manipulations we may eliminate ${\cal G}_k^{(1)}$ by
integration and obtain
\begin{equation}
  \omega^{(1)}\sum_\beta J_\beta |W_\beta^{(0)}|^2=
  \sum_k\int_{a_k}^{b_k}
  \left({{4}\over{{\cal I}_kr^4\Omega^3}}\right)
  \left[-\omega^{(0)}-
      \left({{\kappa^2-\Omega^2}\over{\Omega^2}}\right)
  {{\Omega}\over{2}} + i\alpha\Omega \right]|{\cal G}_k^{(0)}|^2
  \,2\pi r\,dr.
\label{omega(1)}
\end{equation}
Since ${\cal G}_k^{(0)}$ does not depend on $\alpha$ but is
proportional to $\Sigma$, this shows that nearly rigid modes are
damped at a rate ${\rm Im}(\omega^{(1)})$ proportional to
$\alpha/\epsilon^2$ (for a disk and planetary system of given
dimensions).  The imaginary part of equation (\ref{omega(1)}) is
  a special limit of the conservation law derived in Appendix~A.

It is important to note that the self-gravitational term in equation
(\ref{iomegaw}) has no effect on the results we have derived.  The
reason for this is that, if a disk behaves nearly rigidly, {\it the
  self-precession of the disk is negligibly small\/} even if the mass
of the disk exceeds that of the planets.

\section{Mean-motion resonances}

\label{Mean motion}

So far we have treated only secular interactions between the planets
and disks.  However, a planet also interacts with a disk through
mean-motion resonances (Lindblad resonances and vertical resonances).
The launching of waves at these locations exerts a torque on the
planet.  If the planet's orbit is inclined with respect to the
resonant annulus of the disk, this inclination will evolve in time.

Borderies, Goldreich, \& Tremaine (1984, hereafter BGT) calculated the
rate of change of inclination of a satellite due to Lindblad and
vertical resonances.  They showed that the inclination typically grows
exponentially at a rate that depends on the strengths of the
resonances present.  They assumed implicitly that the launched waves
are damped before reaching an edge of the disk from which they might
otherwise reflect.  In the case of a gas disk, such waves can damp by
means of viscous dissipation, radiative damping (Cassen \& Woolum
1996), or wave channeling (Ogilvie \& Lubow 1999).  For narrow gaseous
rings between planets, wave damping may well not take place.  Instead,
reflection from the ring edges may occur, which could greatly reduce
the effects of the mean-motion resonances.  For simplicity, we will
also assume that the waves are damped very close to the resonances and
transfer their angular momentum to the disk there.

To incorporate the effect of mean-motion resonances, we modify our
equations as follows.  (For notational simplicity we here consider
only a single planet and a single disk.)  The formulae of BGT can be
interpreted as giving the horizontal component of the resonant torque
between the planet and the disk.  We use this to add terms to the
planet's angular momentum equation, in the form
\begin{equation}
  m_{{\rm p}}r_{{\rm p}}^2\Omega_{{\rm p}}{{dW_{{\rm p}}}\over{dt}}=
  \cdots+2\pi\sum_j{{Gm_{\rm p}^2}\over{M_*}}s_{{\rm r}j}
  \Sigma r(W_{\rm p}-W)\bigg|_{r=r_{{\rm r}j}},
\end{equation}
where $r_{{\rm r}j}$ is the radius of the $j$th resonance, and
$s_{{\rm r}j}$ is the dimensionless strength of the resonance.
Similarly, we modify the disk's angular momentum equation to
\begin{equation}
  \Sigma r^2\Omega{{\partial W}\over{\partial t}}=
  \cdots+\sum_j{{Gm_{\rm p}^2}\over{M_*}}s_{{\rm r}j}
  \Sigma(W-W_{\rm p})\,\delta(r-r_{{\rm r}j}).
  \label{modified}
\end{equation}
As noted above, this assumes that the waves transfer their angular
momentum to the disk very close to the resonances.  Provided that the
disk is not very cold, the precise location of the wave damping is
unimportant because the torque is then communicated through the disk
over a large radial extent.

The radii and strengths of the resonances are as follows (BGT); here
we have neglected the shifts of the resonant radii caused by
precession.  Inner vertical resonance ($m\ge2$):
\begin{equation}
  x={{r_{\rm r}}\over{r_{\rm p}}}=\left({{m-1}\over{m+1}}\right)^{2/3},
  \qquad
  s_{\rm r}={{\pi x^4}\over{24(m-1)}}\left[b_{3/2}^{(m)}(x)\right]^2.
\end{equation}
Inner Lindblad resonance ($m\ge2$):
\begin{equation}
  x={{r_{\rm r}}\over{r_{\rm p}}}=\left({{m-1}\over{m}}\right)^{2/3},
  \qquad
  s_{\rm r}=-{{m\pi x^2}\over{12(m-1)}}
  \left[\left(2m+x{{d}\over{dx}}\right)b_{1/2}^{(m)}(x)\right]^2.
\end{equation}
Outer vertical resonance ($m\ge2$):
\begin{equation}
  x={{r_{\rm p}}\over{r_{\rm r}}}=\left({{m-1}\over{m+1}}\right)^{2/3},
  \qquad
  s_{\rm r}={{\pi x^2}\over{24(m+1)}}\left[b_{3/2}^{(m)}(x)\right]^2.
\end{equation}
Outer Lindblad resonance ($m\ge1$):
\begin{equation}
  x={{r_{\rm p}}\over{r_{\rm r}}}=\left({{m}\over{m+1}}\right)^{2/3},
  \qquad
  s_{\rm r}={{m\pi}\over{12(m+1)}}
  \left[\left(2m+1+x{{d}\over{dx}}\right)b_{1/2}^{(m)}(x)\right]^2.
\end{equation}
These quantities are listed in Tables~1 and 2 for all resonances
satisfying $|r_{\rm r}-r_{\rm p}|>0.1r_{\rm p}$.  Note that the inner
Lindblad resonances act in the opposite sense to the other resonances.

\placetable{tab1}
\placetable{tab2}

Although equation (\ref{modified}) formally requires the disk tilt $W$
to have a cusp at each resonance, these cusps will be very weak in a
disk with good radial communication.  For a disk that responds nearly
rigidly, the effect of the mean-motion resonances is to add a small
imaginary part to the planet-disk coupling coefficient $C^{\rm pd}$:
\begin{equation}
  C^{\rm pd}\mapsto C^{\rm pd}+2\pi i\sum_j{{Gm_{\rm p}^2}\over{M_*}}
  s_{{\rm r}j}\Sigma_{{\rm r}j}r_{{\rm r}j}.
  \label{imc}
\end{equation}
The sum over resonances is always positive for an exterior disk but
usually negative for an interior disk.  According to equation
(\ref{rigid_sum}), positive sums of this kind tend to cause growth of
the non-trivial normal modes.  This must compete, however, with the
viscous decay implied by equation (\ref{omega(1)}).

\section{Model for disk and numerical method}
\label{Standard model}

We adopt a simplified model for the solar nebula.  The disk model is
similar to that described in Lubow \& Ogilvie (2000).  It is
characterized by an angular semi-thickness $H/r=\epsilon={\rm
  constant}$, and a surface density $\Sigma=\Sigma_0fr^{-3/2}$, where
$\Sigma_0$ is a constant and $f(r)$ a function that is close to unity
in most of the disk, but tapers linearly to zero at the edges
(including gaps around planets).  The width of the tapers is equal to
the local value of $H$.  The vertical structure is that of a polytrope
of index $n$, which gives ${\cal I} = \Sigma H^2/(2n+3)$ (as defined
in eq. [\ref{calI}]) if $H$ is the true semi-thickness of the
polytrope.  The basic parameters of the model are then $\epsilon$,
$\alpha$, and $n$, together with the dimensions and mass of the disk.
We normalize the surface density by quoting a nominal disk mass,
\begin{equation}
  m_{\rm disk}=\int_0^{r_{\rm out}}\Sigma_0r^{-3/2}\,2\pi r\,dr=
  4\pi\Sigma_0r_{\rm out}^{1/2},
\end{equation}
where $r_{\rm out}$ is the outermost radius.  The true disk mass is
slightly less than $m_{\rm disk}$ owing to the various edges.

We consider the nebula in the presence of a Jupiter-like planet.  As
standard parameters, we adopt $m_{\rm p}=0.001M_*$, $m_{\rm disk} =
0.01 M_{*}$, $\epsilon = 0.075$, $\alpha = 0.005$, and $n=3/2$.  The
planet creates a gap in the nebula, leading to two separate disks
whose parameters are taken to be $a_1=0.1r_{\rm p}$, $b_1=(1-g)r_{\rm
  p}$, $a_2=(1+g)r_{\rm p}$, $b_2=20r_{\rm p}$.  The standard relative
half-width of the gap is taken to be $g=0.2$.  This situation may be
compared with Jupiter in a solar nebula extending out to about 100 AU.
The effect of varying the important parameters will be considered
below.


Numerical solutions to the equations of Section~\ref{Normal modes} are
obtained by discretizing the disk, converting the integrals into sums,
and converting derivatives into centered finite differences, with
$W_{{\rm d} k}$ defined on a set of $n_{\rm r}$ rings for each disk,
and ${\cal G}_k$ on the boundaries between neighboring rings.
Planet-disk and disk-disk interactions are computed with softened
kernels $K_m$, as defined by equation (\ref{ksoft}).  The smoothing
length $h=\max(H_1,H_2)$ is taken to be the maximum of the
semi-thicknesses of the two rings concerned.  Equation
(\ref{modified}) is implemented by identifying the ring in which each
resonance falls, and representing the delta function as $1/\delta r$,
where $\delta r$ is the width of the ring.  The problem then reduces
to a generalized eigenvalue problem involving a non-Hermitian matrix
of dimension $n_{\rm p}+n_{\rm d}(2n_{\rm r}-1)$.  This is solved
numerically using the Fortran Sun Performance Library routine ZGEGV on
a Sun Ultra 10, using $n_{\rm r}=200$.  Numerical results are quoted
in units such that $G=M_*=r_{\rm p}=1$.  The modes are normalized such
that $W_{\rm p}=1$.

\section{Numerical results for Jupiter}
\label{Numerical results}

We first consider a disk with standard parameters, except that it
extends only to $b_2=2$ instead of $b_2=20$.  [Accordingly, we set
$m_{\rm disk}=(0.1)^{1/2}0.01$ in this case to fix the same
normalization of surface density as in the standard model.  The true
mass of the disk is only $0.00179$.]  This case is easier to
understand because the disk is able to maintain good radial
communication.

Only two of the non-trivial modes are of interest.  The first mode
(`mode 1') has a precession frequency of $-7.02\times10^{-4}$.  The
inner disk has a nearly rigid tilt $W\approx-20$ and the outer disk
$W\approx+20$.  If the mean-motion resonances are neglected, this mode
has a viscous decay rate of $4.0\times10^{-7}$; when they are
included, the mode acquires a net growth rate of $1.0\times10^{-6}$.
[This mode corresponds to equation (\ref{mode1}) in rigid disk tilt
theory.]

The second mode (`mode 2') has a precession frequency of
$-1.46\times10^{-3}$.  The inner disk has a nearly rigid tilt
$W\approx-0.64$ and the outer disk $W\approx-0.75$.  If the
mean-motion resonances are neglected, this mode has a viscous decay
rate of $7.1\times10^{-7}$; when they are included, the mode acquires
a net growth rate of $6.6\times10^{-6}$.

The remaining non-trivial modes all have significantly larger
precession rates and damping rates (with or without mean-motion
resonances).  They are the proper bending modes of the disks, modified
by gravitational coupling to the other components of the system.

When we proceed to the standard model, the numerical solution becomes
significantly more complicated.  The outer disk extending to $b_2=20$
is not able to maintain good radial communication for typical
precession rates.  The precession frequencies estimated from the
nearly rigid theory are comparable to the frequencies of global
bending modes in the outer disk.  As a result, the nearly rigid theory
is no longer accurate for the outer disk.

To simplify the presentation of results, we focus on three modes that
are of greatest interest in that they involve the least warping, and
have the smallest precession frequencies and damping rates.  `Mode 1'
and `mode 2' are related to the nearly rigid modes mentioned above,
while `mode 3' derives from a bending mode of the outer disk that
interferes with the other two when the outer disk fails to behave
rigidly. The structures of the three modes are plotted in Fig.~1.  The
modes include the effects of mean-motion resonances.

In Figs~2 and 3 we plot the precession rates and decay rates of these
three modes for the standard model, as we vary each of the parameters
$\alpha$, $\epsilon$, $m_{\rm disk}$, and $g$ about its standard
value. Notice that the variation of decay rate with some parameters is
non-monotonic. In particular, the decay rate of mode 1 peaks at
$m_{\rm disk}\approx0.02$.

In Appendix~A, we derive an expression for the local decay rate of a
warped disk due to dissipation.  In Fig.~4 we plot for mode 1 the decay
rate, as given by equation (\ref{gamma}). Notice that the outer disk
dissipation dominates over the inner disk dissipation. Although the
dissipation peaks near the planet, it is broadly distributed
throughout the outer disk. In the outer disk,
the warp $|r dW/dr|$ is 
greatest near the disk radial midpoint and is broadly distributed. 
Quantity $|r d W/dr|$ must vanish
at the inner and outer disk edges, due to boundary condition
(\ref{bcs}). The reason that the peak in Fig.~4 in 
the outer disk occurs near the
planet is that the magnitudes of various base-state quantities, such as
$\Sigma$ and $\Omega$, are largest there.

In summary, an inclined planet gives rise to large-scale warps.  The
damping effects of secular interactions dominate over tilt excitation
by mean-motion resonances in our model system, provided $\alpha \ga
0.001$.  For the standard model, the decay time of the longest lived
mode, mode 1, is about $2.7 \times 10^5$ yr.  Although this time-scale
is short compared to typical disk lifetimes of several million years,
it is comparable to or longer than the expected planetary migration
time-scale at this radius (Lin, Bodenheimer, \& Richardson 1996; Ward
1997).

\section{Simple estimates}

We compare the numerical results with the nearly rigid theory of
Section \ref{Nearly rigid response} for the standard disk model.  To
do this, we carry out some rough estimates and drop all factors of
order unity.  Consistent with our standard model, we assume below that
the mass of the planet is less than, or comparable to, the masses of
the two disks, and also that the angular momentum of the outer disk is
greater than, or comparable to, the angular momenta of the planet and
inner disk.

The coupling coefficients are dominated by the parts of the disk
closest to the planet, where $|r-r_{\rm p}|/r_{\rm p}\sim g$.  Here
the kernel may be estimated as $K\sim1/(g^2r)$.  Thus
\begin{equation}
  C^{\rm pd}\sim{{1}\over{g}}Gm_{\rm p}\Sigma r
\end{equation}
and
\begin{equation}
  C^{\rm dd}\sim G\Sigma^2r^3.
\end{equation}
Here $\Sigma$ is a typical surface density near the planet.  The
precession rate of a mode can be estimated from either equation
(\ref{mode1}) or equation (\ref{mode2}) as
\begin{equation}
  \omega\sim-{{G\Sigma}\over{gr\Omega}}.
\end{equation}

We then estimate ${\cal G}$ from equation (\ref{iomegaw}) as
\begin{equation}
  {\cal G}\sim i\omega\Sigma r^4\Omega W.
\end{equation}
We apply these estimates to equation (\ref{omega(1)}) to obtain an
estimate of the viscous damping rate,
\begin{equation}
  {\rm Im}(\omega)\sim-{{\alpha}\over{\epsilon^2}}{{1}\over{g^2}}
  {{(\Sigma r^2)^2}\over{M_*^2}}\Omega.
\end{equation}

When mean-motion resonances are taken into account, the planet-disk
coupling coefficient acquires a small imaginary part, as in equation
(\ref{imc}).  The sum over resonances can be estimated using arguments
adapted from BGT.  For $g\ll1$, the number of resonances in the disk
scales as $m\sim1/g$.  The strengths of the Lindblad resonances, which
dominate here, scale as $s_{\rm r}\sim m^2$.  Thus
\begin{equation}
  {\rm Im}(C^{\rm pd})\sim{{1}\over{g^3}}
  {{Gm_{\rm p}^2\Sigma r}\over{M_*}}.
\end{equation}
>From equation (\ref{mode1}) this would provide a resonant growth rate
\begin{equation}
  {\rm Im}(\omega)\sim{{1}\over{g^3}}
  {{m_{\rm p}\Sigma r^2}\over{M_*^2}}\Omega.
\end{equation}
The ratio of viscous decay to resonant growth can then be estimated as
\begin{equation}
  {{\rm decay}\over{\rm growth}}\sim
  \left({{\alpha}\over{\epsilon^2}}\right)
  \left({g{\Sigma r^2}\over{m_{\rm p}}}\right).
\end{equation}
The first factor depends on the viscosity and temperature of the disk.
The second is related to the ratio of the disk mass within (say) two
gap widths of the planet to the mass of the planet itself. For our
standard parameters, these first factor is close to unity, while the
second is somewhat less than unity.  This favors tilt growth,
provided that the disk indeed behaves nearly rigidly.

\section{Connection with WKB bending wave theory}

\label{WKB bending wave}

Disks subject to forcing by a misaligned companion have been studied
in the context of planetary rings, but with strikingly different
results. Within Saturn's A ring, tightly wrapped waves are launched
from a mean-motion vertical resonance by a misaligned satellite, as
described by Shu, Cuzzi, \& Lissauer (1983).  Furthermore, the torque
carried by the waves is independent of $\alpha$ and $\epsilon$, unlike
the present case.

As emphasized earlier, the difference lies in the fact that the waves
here are of low frequency such that the parameter $\omega r/c_{\rm s}$
is typically less than unity for protostellar disks with secular
resonances.  For the mean-motion vertical resonances in planetary
rings, this quantity is much greater than unity.

Nonetheless, a formal connection between the equations used in this
paper and those used for planetary rings can be made by considering an
artificially cold protostellar disk, such that $\omega r/c_{\rm s}$ is
much greater than unity.  Furthermore, we consider the case that the
disk mass is much greater than the planetary mass.  Appendix~B
outlines the derivation of the dynamical equations in the WKB limit.
In that limit we obtain
\begin{equation}
  \label{WKB}
  D(r)W-2\pi i\sigma_k G\Sigma\frac{dW}{dr}
  +\left(\frac{\cal I}{2\Sigma}\right)\frac{\Omega^3}
  {[\omega+(\kappa^2-\Omega^2)/(2\Omega)-i\alpha\Omega]}
  \frac{d^2W}{dr^2}
  \approx\frac{2G m_{\rm p}}{r^2} K(r_{\rm p},r) W_{\rm p},
\end{equation}
where $\sigma_k = \pm 1$ with the sign depending on whether the waves are
trailing or leading, and
\begin{equation}
  \label{res}
  D(r)=2\omega\Omega+\frac{2G m_{\rm p}}{r^2} K(r_{\rm p},r).
\end{equation}
For $\omega \ll \Omega$, we can re-express $D(r)$
in the standard form
\begin{equation}
  \label{D}
  D(r) = \mu'^2 - (\Omega' - \omega)^2,
\end{equation}
where
\begin{equation}
  \mu'^2 = {{\partial^2 \Phi'} \over {\partial z^2}}\Bigg|_{z=0},\qquad
  r\Omega'^2={{\partial \Phi'}\over{\partial r}}\Bigg|_{z=0}.
\end{equation}
Potential $\Phi'$ is due to the star and axisymmetric contributions of
the planet, {\it but does not include contributions due to the disk}.
This equation is identical in form to equation (20) of Shu et al.
(1983) for $m=1$, apart from the pressure term, the third term in the
equation (\ref{WKB}).

The pressure term arises because the warping of the disk introduces
horizontal pressure gradients.  The resulting shearing horizontal
motions give rise to the hydrodynamic torque $2\pi\bG$, which is the
dominant factor in the propagation of bending waves in protostellar
disks.  The warping is described by the $n=0$
  mode in the notation Lubow and Pringle (1993). 
The dispersion relation for adiabatic oscillations of this
mode in an exactly Keplerian disk, whose unperturbed state is
vertically isothermal, follows from equation (54) of Lubow and Pringle
(1993) with dimensionless frequencies $F = (\Omega - \omega)/\Omega
\approx 1$ and $\kappa = 1$.
In the low-frequency
regime ($|\omega| \ll \Omega$), one obtains then that 
 $\omega = \pm k H \Omega
/2$,  independent of the exponent for adiabatic compression $\gamma$.
Although the form of the dispersion relation is reminiscent of that for
a compressive acoustic mode, this mode is largely incompressible, as
suggested by the lack of dependence on $\gamma$ in the dispersion
relation.  The same dispersion relation (as was obtained by Papaloizou
\& Lin 1995) follows from equation (\ref{WKB}), with $W$ varying as
$\exp{(ikr)}$, and ${\cal I} = (c_{\rm s}/\Omega)^2 \Sigma$, and by taking
$G=0$ (ignoring gravity), $\kappa = \Omega$, and $\alpha = 0$.

For this limit where the planet mass is small compared to the disk
mass, the resonance condition $D(r) = 0$ becomes
\begin{equation}
\label{rescon}
\frac{m_{\rm p}}{m_{\rm d}} \frac{K(r, r_{\rm p})}{\langle K \rangle}
\left(\frac{r_{\rm p}}{r}\right)^{1/2} = 1
\end{equation}
where $\langle K \rangle$ denotes the mass weighted average of $K$
over the disk (see Appendix~B for more details).  For the standard
disk model, the resonance condition is satisfied at radial distances
of about 0.7 and 1.3 within the inner and outer disks respectively.

This system contains a secular particle resonance where the precession
rate of a particle matches the precession rate of the disk-planet
system.  A free particle (e.g. an asteroid) will precess due to the
gravitational effects of both the disk and the planet.  That is, a
free particle resonance will occur when $D(r) = 0$ in equation
(\ref{D}), but $\Phi'$ now has an additional contribution due to the
disk. The particle resonance location is therefore not the same as the
disk resonance location. In other words,
in locating a particle secular resonance, the gravity of
the gas disk must
be included.  Particles oscillating vertically in a
non-oscillating gas
disk have an additional vertical restoring force that alters their
precession
frequency and shifts their resonance site. This is a special property of the $m=1$
bending modes (tilt modes), which is a consequence of the fact that
there is no self-precession in a rigidly tilted disk. Near resonance,
a tilt mode essentially behaves rigidly, due to its locally long
wavelength.

In the case of a planetesimal (particle) disk, the nature of the
pressure and viscous interaction may well differ from that of a
gaseous disk. On the other hand, the secular resonance condition given
by equation (\ref{rescon}) likely applies to a low-mass planet
interacting with a cold planetesimal disk.

\section{Planet in central hole}

Current models for the early evolution of close-orbiting extra-solar
planets, such as 51 Peg b (Mayor \& Queloz 1995) with an orbital
radius of about 0.05 AU, involve the planet migrating into a central
hole in the disk (Lin et al. 1996). The planet is envisioned to
migrate into the hole by disk tidal forces until its 2:1 outer
Lindblad resonance lies just inside the inner edge of the disk. The
planet would then remain at this radius, without further migration.

We consider the tilt stability involving both secular interactions and
mean-motion resonances of such a planet-disk system, based on the
standard model parameters in Section \ref{Standard model}. In this
case, there is only an outer disk of mass 0.01 $M_{*}$, which extends
from just outside the planet's 2:1 outer Lindblad resonance $1.31
r_{\rm p}$ (0.066 AU) to outer radius 2000 $r_{\rm p}$ (100 AU). Other
disk parameters are the same as in the standard model ($\epsilon =
0.075$, $\alpha = 0.005$, and $n=3/2$).  The only mean-motion
resonance present in the disk is the rather weak 1:3 resonance (2.08
$r_{\rm p}$), which acts to increase inclination.

Fig.~5 shows the eigenfunction of the lowest mode of the system.  The
eigenfrequency is $-5.11\times 10^{-7}+4.6\times10^{-8}i$ in units of
$\Omega_{\rm p}$, corresponding to a decay time-scale of only $3.9
\times 10^4$ yr.  The inner part of the disk interacts nearly rigidly
with the planet, owing to good radial communication $\omega r_{\rm
  p}/c_{\rm s} \ll 1$ (see lower panel of Fig.~5).  However, the outer
part of the disk is not in good communication ($\omega r/c_{\rm s}$ is
of order unity), resulting in a warp and dissipation.

\section{Disk and two planets}
\label{Disk and two planets}

We consider here the case of Jupiter and Saturn interacting with the
portion of the solar nebula that lies interior to the orbit of
Jupiter, using the equations of Section \ref{Normal modes}.  This case
is of interest because the classical $\nu_{16}$ nodal secular
resonance lies within the disk.  We ignore the effects of mean-motion
resonances.

The rigid tilt model involving three objects (here the two planets
plus disk) was considered at the end of Section \ref{Rigid response}.
As is mentioned there, a secular resonance of the two planets with the
disk is possible, if the disk is low in mass compared with the
planets.  If the disk mass is greater than the planet masses, then the
planets individually interact with the disk in a manner similar to the
case in Section \ref{Numerical results}.

We have computed the modes of the system numerically.  We consider
Jupiter (planet 1) and Saturn (planet 2) in circular orbits with the
present values of mass ($m_{{\rm p}1}=0.0009545$, $m_{{\rm
    p}2}=0.0002858$) and semi-major axis ($r_{{\rm p}2}=1.833$).  A
test particle would experience nodal secular resonances at semi-major
axes $0.381$ and $2.39$, corresponding to $1.98$ AU and $12.4$ AU
(Murray \& Dermott 1999).  We consider a partially depleted solar
nebula interior to Jupiter's orbit.  Our standard model has $m_{\rm
  disk}=0.001$, $\epsilon=0.075$, $\alpha=0.005$, $n=3/2$, and disk
inner and outer radii $a_1=0.1$ and $b_1=1-g=0.8$, respectively, in
units of Jupiter's orbital radius.  The inner secular resonance for a
test particle therefore lies inside the disk.

For these parameters, the numerically determined eigenfrequencies of
the two lowest modes are $-2.27\times10^{-4}+6.0\times10^{-9}i$ and
$-7.66\times10^{-4}+5.1\times10^{-7}i$.  The response of the disk is
very nearly rigid and nothing special happens at the location of the
secular particle resonance.  This is to be expected, since $\omega
r/c_{\rm s} \ll 1$ throughout the disk.  The effect of
self-gravitation is negligible even though the mass of the disk is
comparable to that of the planets.

If a low-mass disk is extremely cold, it is unable to establish radial
communication very effectively on secular time-scales.  It cannot
respond nearly rigidly, but instead launches a wave at the location of
the particle resonance.  (The rigid response is essentially the limit
of a wavelike response when the wavelength becomes large compared to
the size of the disk.)  In Fig.~6 we show the wavelike response
generated at the $\nu_{16}$ secular resonance at $2$ AU.  A very thin
disk ($\epsilon=10^{-5}$) is required to see this effect, and even
then the wavelength is not much shorter than the size of the disk (so
a WKB or tight-winding treatment of the wave might not be particularly
accurate).  We set $\alpha=\epsilon=10^{-5}$ so that the wave would be
partially damped on reaching the inner boundary.  We also chose a low
disk mass, $m_{\rm disk}=10^{-6}$, so that the disk has a Toomre
parameter $Q>1$.  The eigenfrequency for this lowest mode is
$-2.35\times10^{-4}+1.6\times10^{-8}i$, corresponding to a damping
time of about $1.1\times10^8$ yr.  The resonantly launched waves damp
inclination faster than those in the warmer disk model, even though
the disk mass is much lower.

\section{Summary}

We have carried out a normal mode analysis of planet-disk systems that
includes the effects of secular interactions and mean-motion
resonances. The planets and disks interact gravitationally, and the
disks communicate internally through gas pressure, self-gravity, and
turbulent viscosity.  Secular interactions of  misaligned planet-disk
systems give rise to global effects in the disks, owing mainly to the
effects of gas pressure. The low frequencies $\omega$ associated with
secular modes allow pressure to communicate effectively over radial
distances in disks that are comparable to (and somewhat greater than)
the planet's orbital radius. Over such distances, disks behave rigidly
with little dissipation. On the other hand, over larger distances
($\sim c_{\rm s}/\omega$), which may arise in a 
continuous disk exterior to the
planet's orbit, the radial communication breaks down and large-scale
warps occur, along with enhanced dissipation (see Figs~1 and~5).  The
protostellar disk response to secular interactions typically falls
between that of a global nearly rigidly tilted disk (see Section
\ref{Nearly rigid response}) and that of a locally launched wave (see
Sections \ref{WKB bending wave} and \ref{Disk and two planets}).

Secular interactions act to decrease inclination by means of turbulent
dissipation of horizontal shearing motions within the warp for a
simple $\alpha$ disk prescription. The dissipation of the warp is
peaked somewhat near the planet, but is broadly distributed throughout
the disk (Fig~4).  Secular interactions are in competition with the
effects of mean-motion resonances (BGT), which can act to increase
inclination.  For standard disk parameters, the secular interactions
involving a single planet
generally dominate and the inclination decays in time. For a young
Jupiter interacting with the solar nebula, tilt decay takes places for
$\alpha \ga 0.001$, with a typical alignment time-scale of order
$10^{5}$ yr (see Fig.~3).  For a close-orbiting planet in the central
hole of the disk, the alignment is more rapid.  These results suggest
that a planet formed within a disk will remain coplanar with the disk,
at least as a consequence of disk interactions.

The classical $\nu_{16}$ particle secular resonance, due to Jupiter
and Saturn, is expected to have resided within the solar nebula.
Again, due to pressure effects, this resonance is greatly broadened in
the gas disk.  Consequently, the nebula responds nearly rigidly, with
some decay in inclination.  A very cold planetesimal disk might
exhibit a genuinely wavelike response (cf. Fig~6).

The analysis in this paper has focused mainly on the effects
of a single planet interacting with its interior and exterior
disks. For a system of a multiple planets that opens
multiple disk gaps, the situation is more complicated
because of reflections that occur at the disk edges associated with the gaps.
Such gaps can decrease the effective size of the disk region 
over which a planet can interact, thereby resulting in a weaker
warp and weaker dissipation.
The hydrodynamic torque cannot propagate across a gap, while
the gravitational torque may at least partially do so.
It is quite possible that much longer lived noncoplanar modes may 
result if the outer part of the nebula is almost disconnected
in this way.

\acknowledgments

This work was supported by NASA grants NAG5-4310 and NAG5-10732, the
STScI visitor program, and the Institute of Astronomy visitor program.
GIO acknowledges the support of Clare College, Cambridge through a
research fellowship and the Royal Society through a University
Research Fellowship.

\appendix

\section{Derivation of the local decay rate of bending disturbances}

Starting from equations (\ref{planet})--(\ref{torque}) we obtain the
conservation law
\begin{equation}
  {{d}\over{dt}}(-L_z)=-\sum_k\int_{a_k}^{b_k}
  {{4\alpha|{\cal G}_k|^2}\over{{\cal I}_kr^4\Omega^2}}2\pi r\,dr,
\end{equation}
where
\begin{equation}
  -L_z=\sum_i{{1}\over{2}}
  m_{{\rm p}i}r_{{\rm p}i}^2\Omega_{{\rm p}i}|W_{{\rm p}i}|^2+
  \sum_k\int_{a_k}^{b_k}\left(
  {{1}\over{2}}\Sigma_kr^2\Omega|W_{{\rm d}k}|^2+
  {{2|{\cal G}_k|^2}\over{{\cal I}_kr^4\Omega^3}}\right)
  2\pi r\,dr.
\end{equation}
The quantity $L_z$ is the vertical angular momentum associated with
the bending disturbance.  When an initially horizontal ring of matter
having orbital angular momentum $J$ is tilted through an angle
$\beta=|W|$, the change in the vertical component of angular momentum
is $J(\cos\beta-1)\approx-{\textstyle{{1}\over{2}}}J|W|^2$ when
$|W|\ll1$.  The term proportional to $|{\cal G}|^2$ represents an
additional angular momentum stored in the bending wave.  For a
short-wavelength (WKB) bending wave in a non-self-gravitating disk,
both terms in the integral for $-L_z$ are equal.  For a nearly rigidly
tilted disc, the first term in the integral dominates.  In the
appropriate limit, this first term agrees with the expression for the
angular momentum of a bending wave given by Bertin \& Mark (1980, eq.
[C12]).

In an inviscid system $L_z$ is conserved.  When viscosity is present,
the positive definite quantity $-L_z$ decays monotonically to zero.
The negative angular momentum of the bending disturbance is
transferred to the disk through viscosity, and causes accretion.  (It
is nevertheless consistent that we neglect the time-dependence of
$\Sigma$ in a linear theory, because the induced accretion rate is
quadratic in $W$.)

We now define a local damping rate $\gamma(r)$ by
\begin{equation}
  \gamma=\left(-{{1}\over{L_z}}\right)
  {{2\alpha |{\cal G}|^2}\over{{\cal I}r^4\Omega^2}} 2\pi r \sum_k(b_k-a_k).
  \label{gamma}
\end{equation}
This has the property that
\begin{equation}
  {{\sum_k\int_{a_k}^{b_k}\gamma\,dr}\over{\sum_k(b_k-a_k)}}=
  -{{1}\over{2}}{{d}\over{dt}}\ln(-L_z).
\end{equation}
In the case of a normal mode, the radial average of $\gamma$ corresponds
to the damping rate ${\rm Im}(\omega)$ of the mode.

\section{Derivation of the bending wave equations in the WKB limit}

In this Appendix we derive equations (\ref{WKB}) and (\ref{rescon}) of
the text.  We consider equations (\ref{wpnm})--(\ref{dwdr}) for a
single planet interacting with a single disk.
\begin{eqnarray}
  i\omega m_{\rm p} r_{\rm p}^2 \Omega_{\rm p} W_{\rm p}&=&
  \int Gm_{\rm p}\Sigma K(r,r_{\rm p})
  i(W -W_{\rm p})\,2\pi r\,dr,
\label{wp-app}\\
  i\omega\Sigma r^2\Omega W&=&
  {{1}\over{r}}{{d{\cal G}}\over{dr}}+
   Gm_{\rm p} \Sigma K(r,r_{\rm p})i
    (W_{\rm p} - W) \nonumber\\
  &&+\int
     G\Sigma \Sigma' K(r,r')i(W'-W)
    \,2\pi r'\,dr',
\label{w-app}
\end{eqnarray}
\begin{equation}
  i\omega{\cal G}+   \left({{\kappa^2-\Omega^2}\over{\Omega^2}}\right)
  {{i\Omega}\over{2}}{\cal G}+\alpha\Omega{\cal G} =
  {{{\cal I} r^3\Omega^3}\over{4}}{{dW}\over{dr}}.
\label{G-app}
\end{equation}

We now apply the WKB approximation for a cold disk, $\omega r \gg
c_{\rm s}$.  In this regime we assume
\begin{equation}
H^{-1} \gg \bigg| \frac{\partial \ln{W}}{\partial r} \bigg| \gg r^{-1}.
\end{equation}
The requirement on $H^{-1}$ is used in the derivation of the warp
equations (see Papaloizou \& Lin 1995; Lubow and Ogilvie 2000).  From
equation (\ref{G-app}), we obtain
\begin{equation}
  {{1}\over{r}}{{d{\cal G}}\over{dr}} \approx
-\frac {i}{4}\frac{ {\cal I} r^2 \Omega^3}
  {[\omega+(\kappa^2-\Omega^2)/(2\Omega)-i\alpha\Omega]}
  \frac{d^2W}{dr^2}.
\label{G-WKB}
\end{equation}
For the self-gravity term, we have that
\begin{equation}
\int
     G\Sigma \Sigma' K(r,r')i(W'-W)
    \,2\pi r'\,dr' \approx \pm \pi G\Sigma^2 r^2 \frac{dW}{dr}.
\label{gr-WKB}
\end{equation}
To obtain equation (\ref{gr-WKB}), we recognize that self-gravity is
dominated by local contributions in the WKB limit and apply the
approximation that $K(r, r') \simeq r/[2 \pi (r'-r)^2]$ for $r' \simeq
r$. The integral is computed by extending $r'$ to the complex plane
and integrating over a closed contour. The contour begins along the
negative real axis, includes a small semicircle around the singularity
at $r' = r$, continues along the real axis, and closes on itself
through a large semicircle. The large semicircle resides in either the
upper or lower half-plane, depending on whether the waves are trailing
or leading. Equation (\ref{gr-WKB}) then follows from the residue
theorem.  This choice of trailing or leading waves determines the sign
of the result.  Substituting equations (\ref{G-WKB}) and
(\ref{gr-WKB}) into equation (\ref{w-app}), we obtain equation
(\ref{WKB}).

Integrating equation (\ref{w-app}) over the disk area and using
equation (\ref{wp-app}), we have that
\begin{equation}
J_{\rm d} \langle W \rangle = - J_{\rm p} W_{\rm p},
\end{equation}
where $\langle W \rangle$ denotes the angular momentum weighted
average of $W$. In the limit that $J_{\rm p} \ll J_{\rm d}$, we expect
that $W$ on the right hand side of equation (\ref{wp-app}) can be
ignored after integration.  It then follows from equation
(\ref{wp-app}) that
\begin{equation}
\omega = - \frac{\int G \Sigma K(r,r_{\rm p}) 2 \pi r\, dr}{r_{\rm p}^2
\Omega_{\rm p}} = - \frac{C^{\rm pd}}{J_{\rm p}}.
\end{equation}
Applying this equation for $\omega$ to the resonance condition $D(r) =
0$, for $D$ defined by equation (\ref{res}), we obtain the resonance
condition (\ref{rescon}).

\begin{deluxetable}{rrrr}
\tablecaption{Mean-motion resonances interior to the planet\label{tab1}}
\tablewidth{0pt}
\tablehead{$r_{\rm r}/r_{\rm p}$&$s_{\rm r}$&\colhead{Type}&\colhead{$m$}}
\startdata
$0.4807$&$0.0133$&IVR&2\\
$0.6300$&$-1.1780$&ILR&2\\
$0.6300$&$0.0698$&IVR&3\\
$0.7114$&$0.1916$&IVR&4\\
$0.7631$&$-3.7521$&ILR&3\\
$0.7631$&$0.4009$&IVR&5\\
$0.7991$&$0.7200$&IVR&6\\
$0.8255$&$-7.6763$&ILR&4\\
$0.8255$&$1.1716$&IVR&7\\
$0.8457$&$1.7780$&IVR&8\\
$0.8618$&$-12.9485$&ILR&5\\
$0.8618$&$2.5617$&IVR&9\\
$0.8748$&$3.5451$&IVR&10\\
$0.8855$&$-19.5683$&ILR&6\\
$0.8855$&$4.7508$&IVR&11\\
$0.8946$&$6.2012$&IVR&12\\
\enddata
\end{deluxetable}

\begin{deluxetable}{rrrr}
\tablecaption{Mean-motion resonances exterior to the planet\label{tab2}}
\tablewidth{0pt}
\tablehead{$r_{\rm r}/r_{\rm p}$&$s_{\rm r}$&\colhead{Type}&\colhead{$m$}}
\startdata
$2.0800$&$0.0191$&OVR&2\\
$1.5874$&$1.4925$&OLR&1\\
$1.5874$&$0.0880$&OVR&3\\
$1.4057$&$0.2272$&OVR&4\\
$1.3104$&$4.3077$&OLR&2\\
$1.3104$&$0.4589$&OVR&5\\
$1.2515$&$0.8055$&OVR&6\\
$1.2114$&$8.4664$&OLR&3\\
$1.2114$&$1.2895$&OVR&7\\
$1.1824$&$1.9334$&OVR&8\\
$1.1604$&$13.9710$&OLR&4\\
$1.1604$&$2.7595$&OVR&9\\
$1.1431$&$3.7904$&OVR&10\\
$1.1292$&$20.8221$&OLR&5\\
$1.1292$&$5.0485$&OVR&11\\
$1.1178$&$6.5564$&OVR&12\\
$1.1082$&$29.0200$&OLR&6\\
$1.1082$&$8.3364$&OVR&13\\
$1.1001$&$10.4110$&OVR&14\\
\enddata
\end{deluxetable}

\begin{figure}
  \centerline{\epsfbox{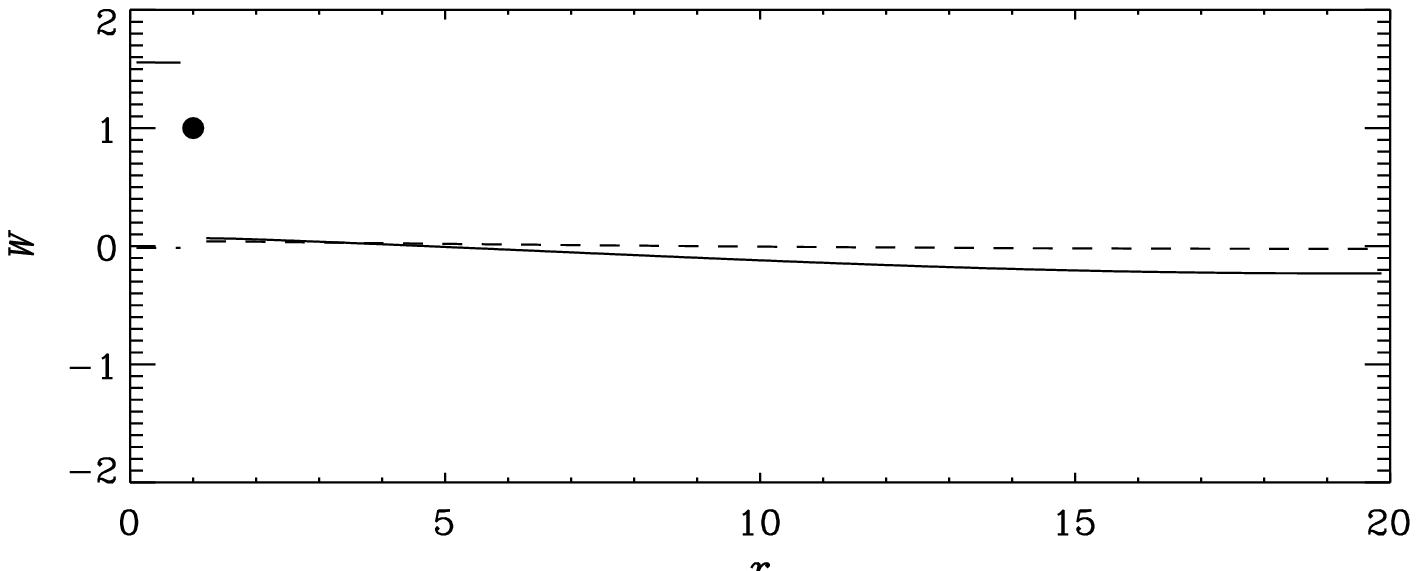}}
  \centerline{\epsfbox{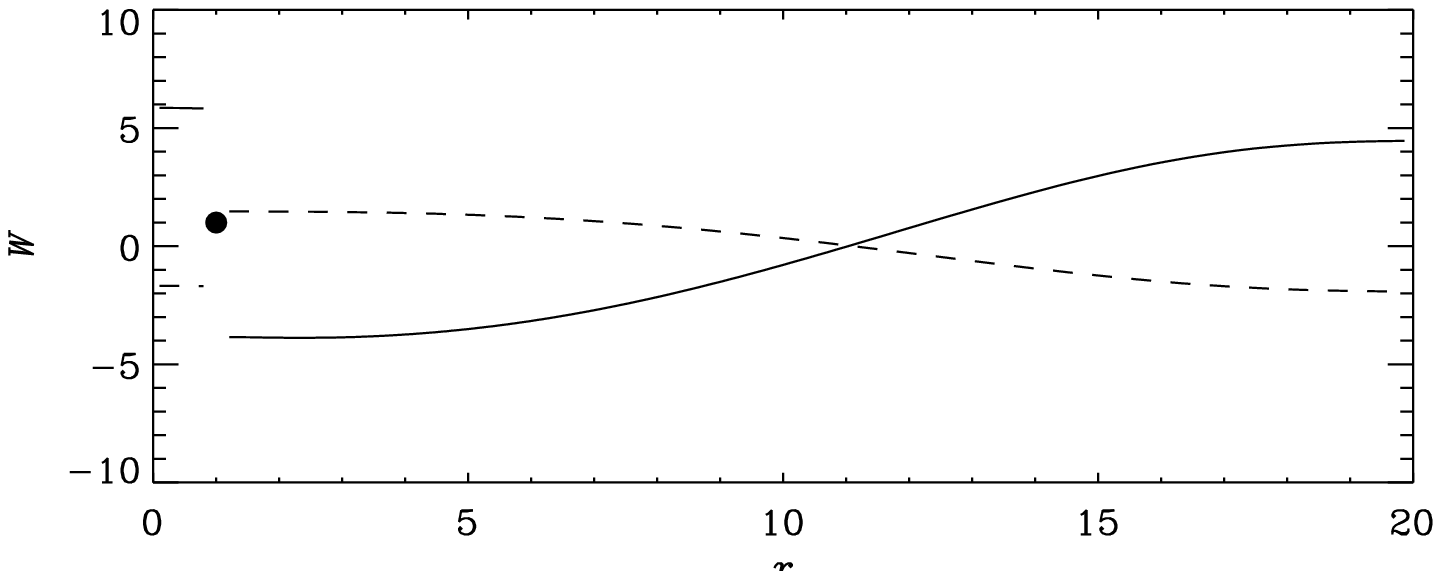}}
  \centerline{\epsfbox{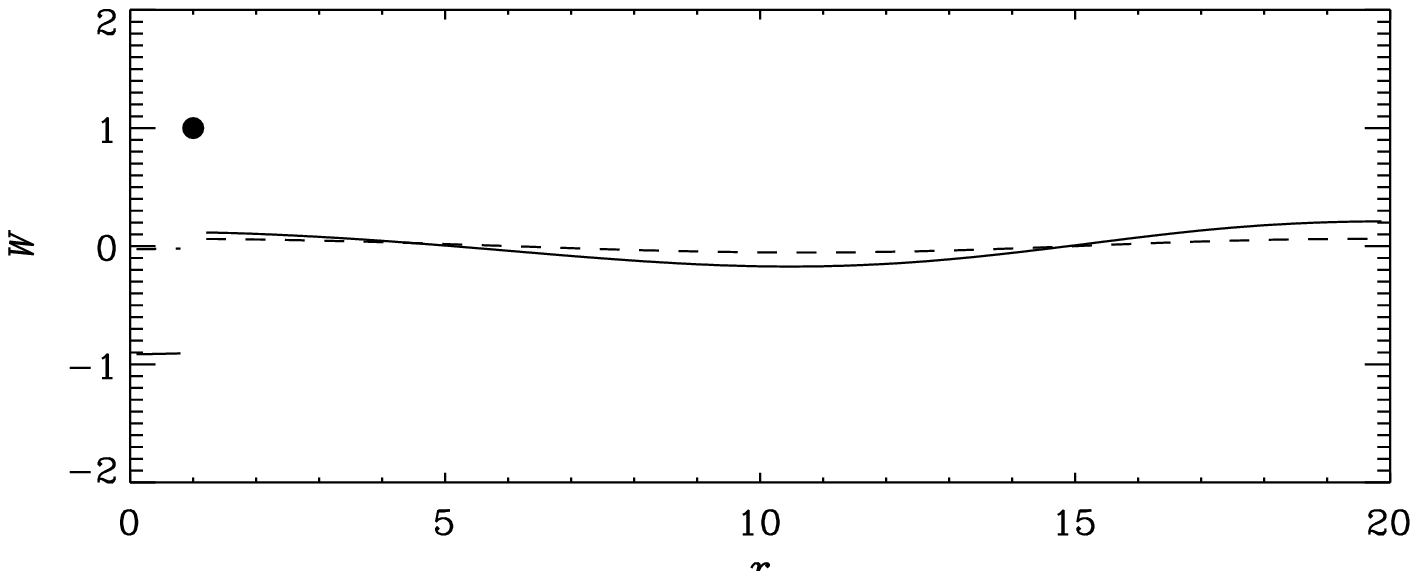}}
  \figcaption[]{Eigenfunctions of the
    tilt variable $W$ for the three lowest modes, with mode 1 plotted
    at the top, mode 2 in the middle, and mode 3 at the bottom. The
    plots are normalized such that the the tilt of the planet (dot) is
    unity at a radius of unity.  The solid curves are for the real
    part of the eigenfunction, while the dashed curves denote the
    imaginary part. For the particular disk model adopted (with
    constant $H/r$), $W(r)$ is also equal to the vertical displacement
    from the midplane in units of the local disk scale-height, if the
    planet resides at one disk scale-height above the midplane.}
\end{figure}

\begin{figure}
  \centerline{\epsfbox{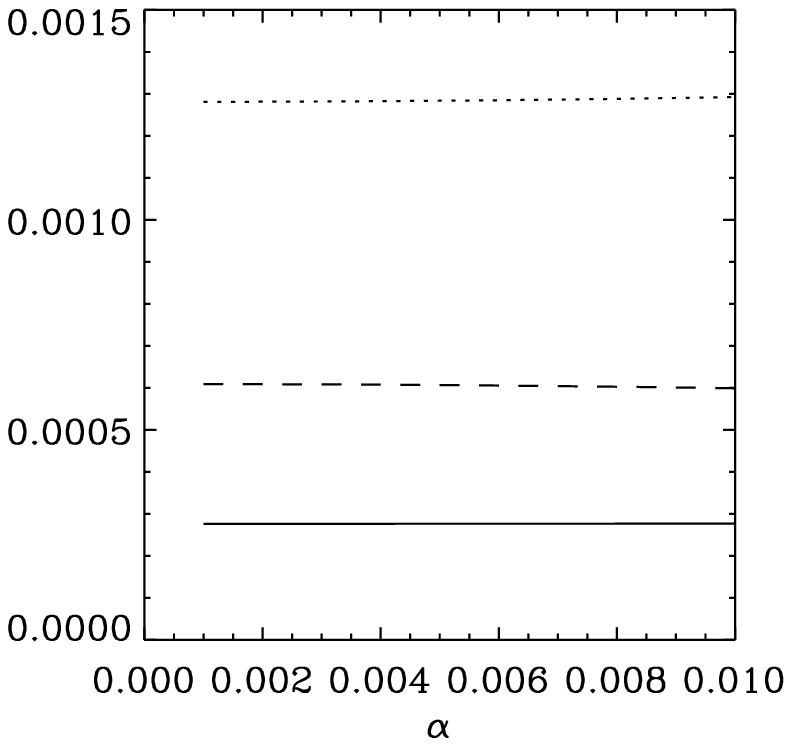}\qquad\epsfbox{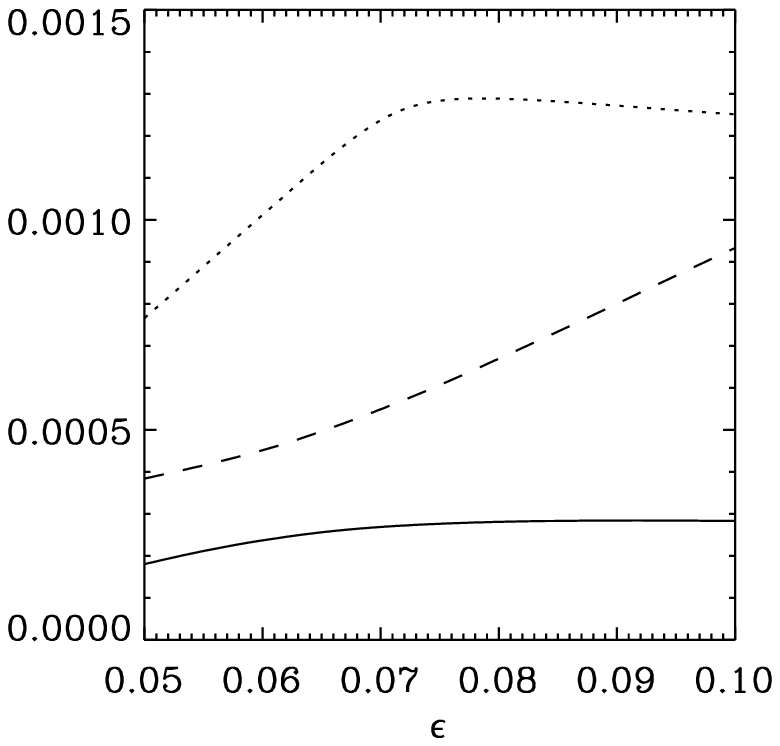}}
  \centerline{\epsfbox{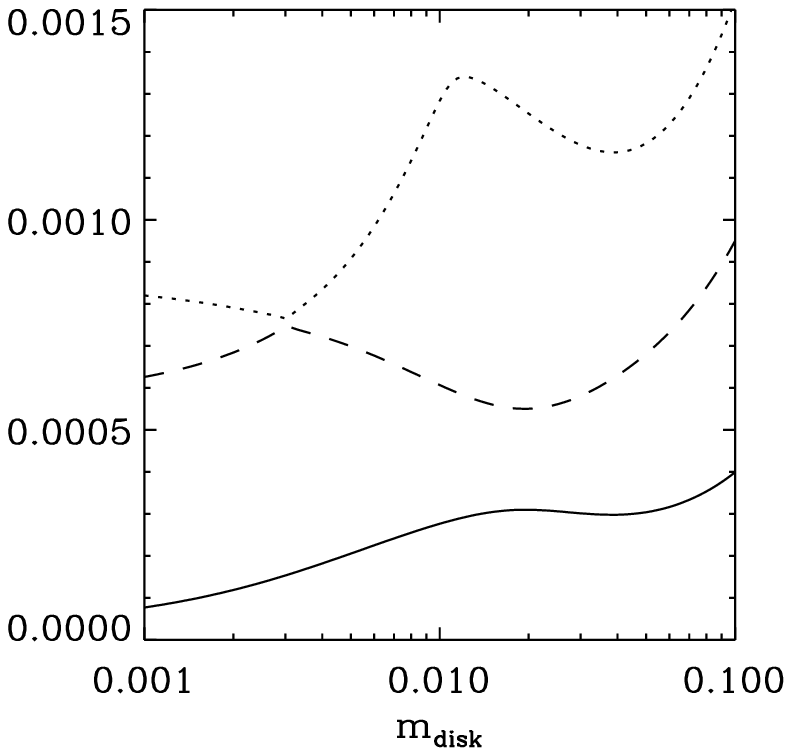}\qquad\epsfbox{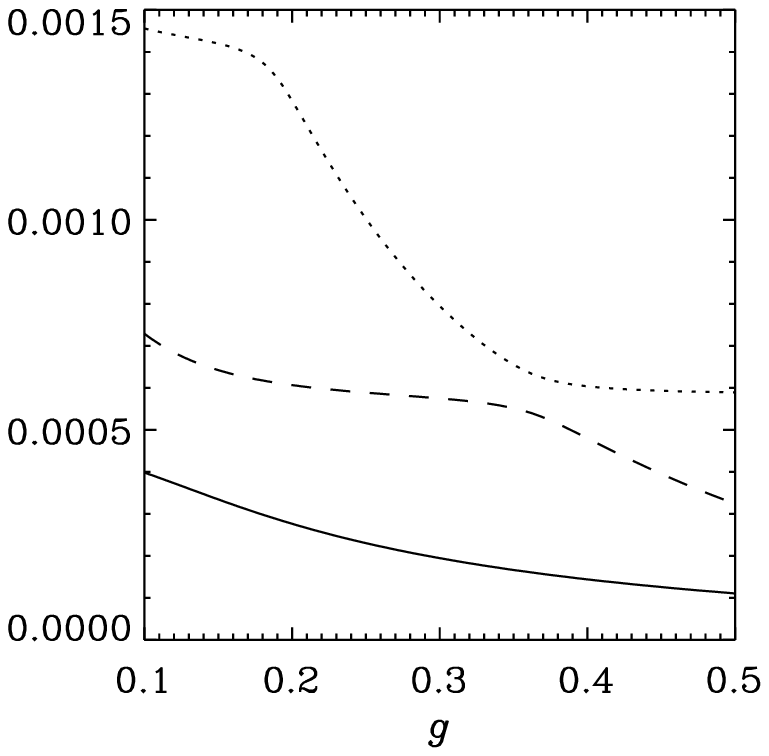}}
  \figcaption[]{Variation of the precession rates of the three lowest
    modes as the parameters $\alpha$, $\epsilon$, $m_{\rm disk}$, and
    $g$ are varied independently about their standard values.  The
    solid, dashed, and dotted lines correspond to modes 1, 2, and 3
    referred to in the text.}
\end{figure}

\begin{figure}
  \centerline{\epsfbox{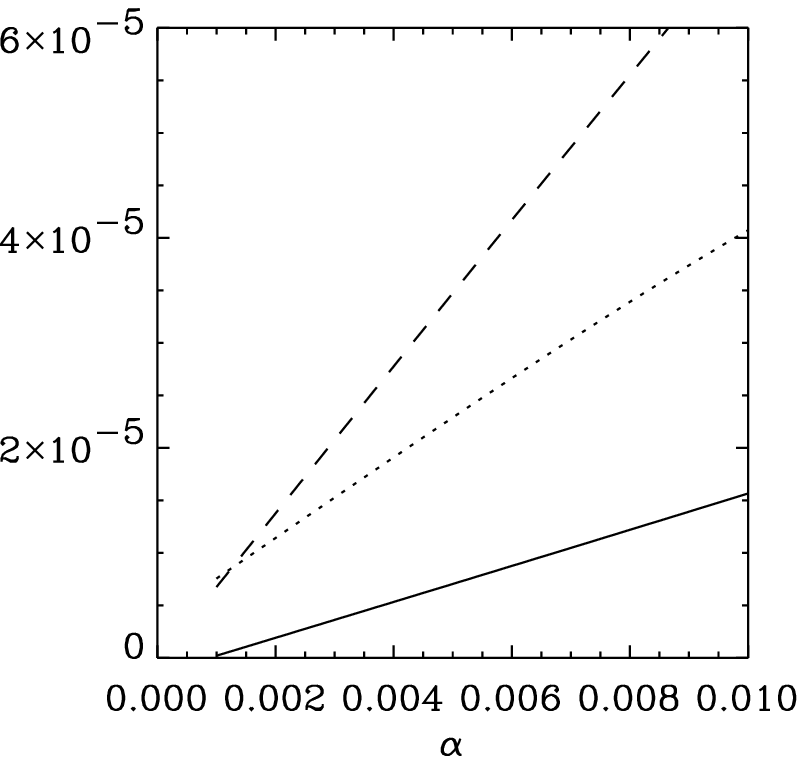}\qquad\epsfbox{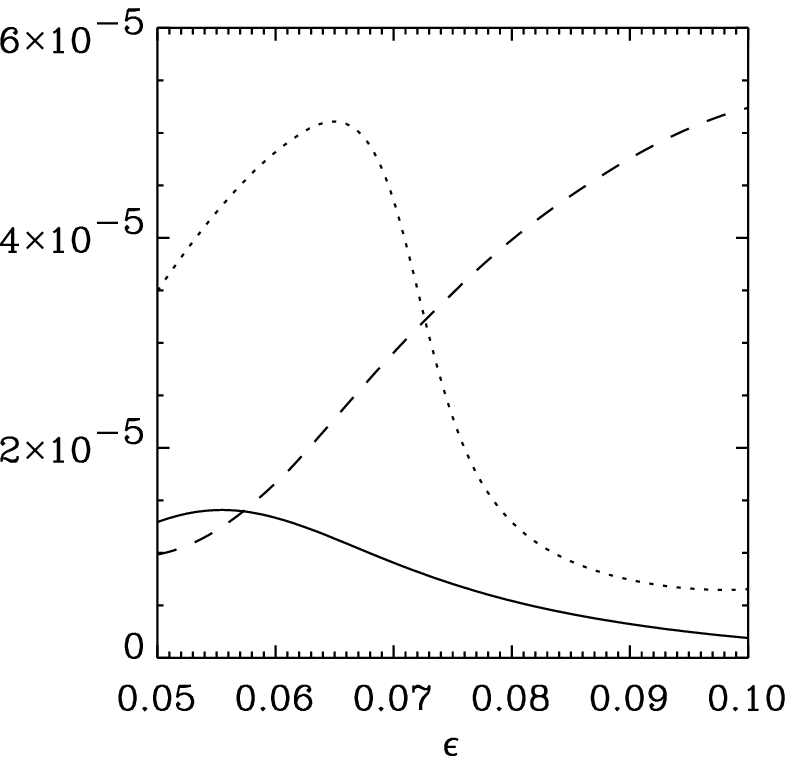}}
  \centerline{\epsfbox{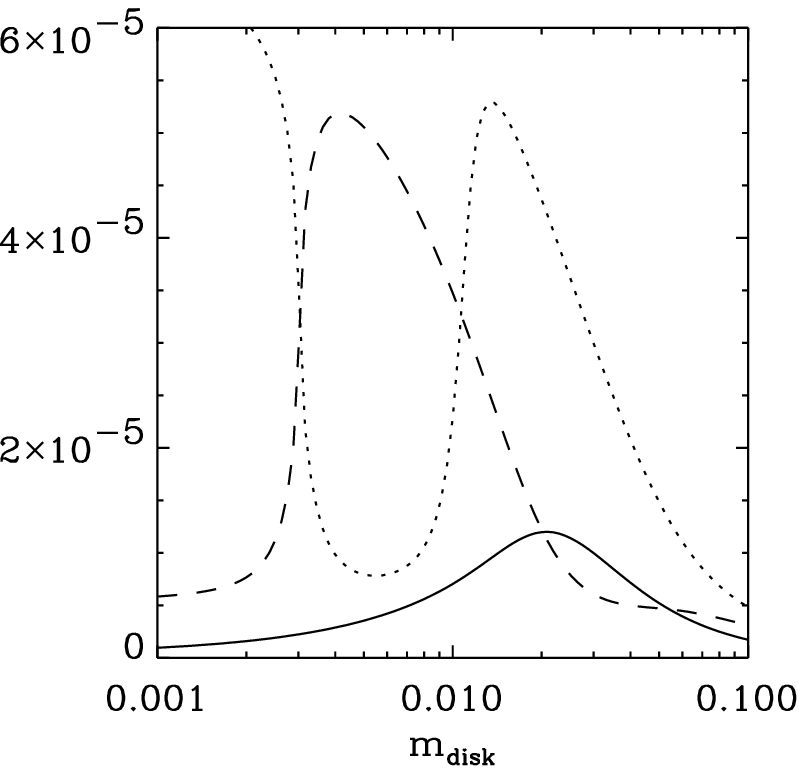}\qquad\epsfbox{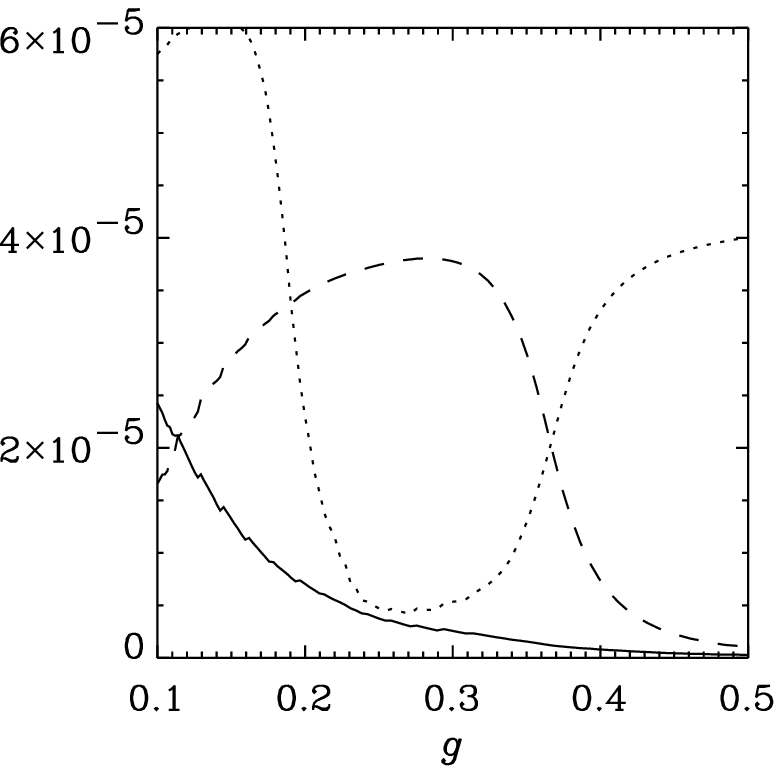}}
  \figcaption[]{Variation of the decay rates. The plot follows the notation
   of Fig.~2.}
\end{figure}

\begin{figure}
  \centerline{\epsfbox{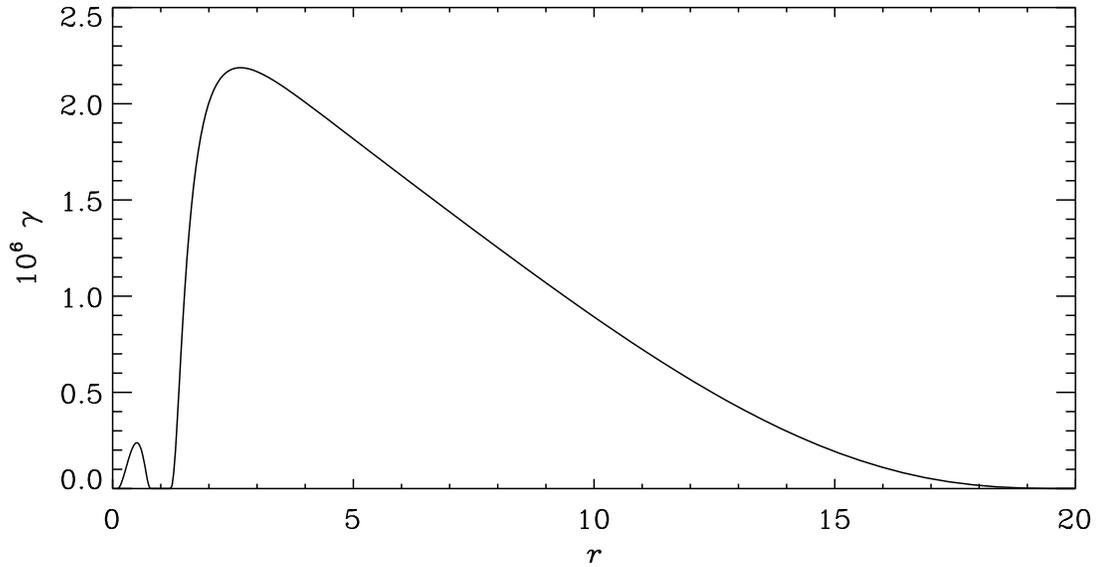}}
  \figcaption[]{ Local decay rate of mode 1, as defined in Appendix~A,
   for the standard model.}
\end{figure}

\begin{figure}
  \centerline{\epsfbox{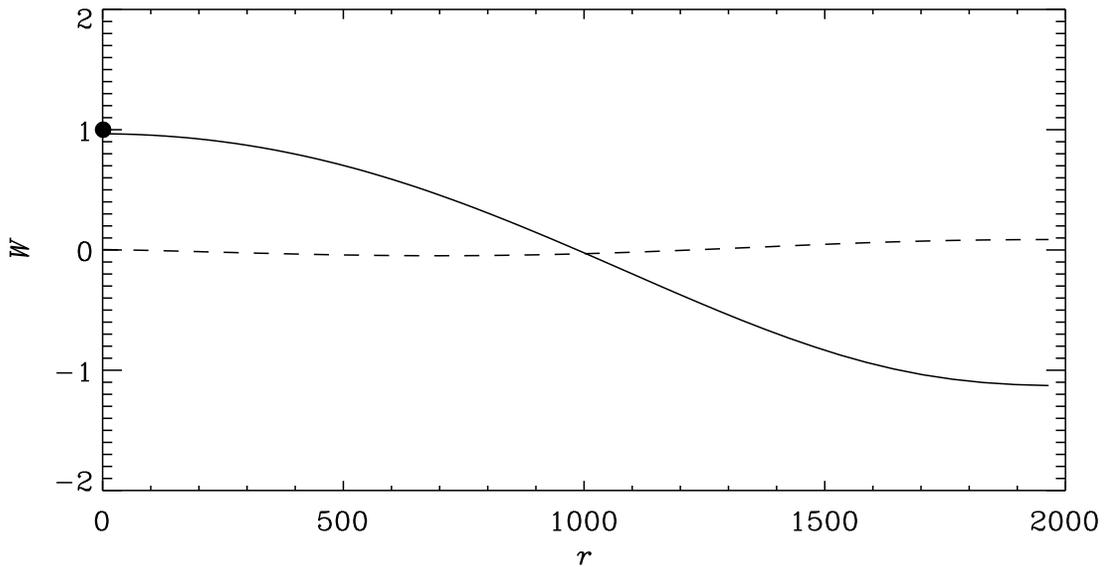}}
  \figcaption[]{Eigenfunction of
    the tilt variable $W$ for the lowest order mode involving a
    close-orbiting planet in the central hole of a disk. The plots
    follow the notation of Fig.~1.}
\end{figure}

\begin{figure}
  \centerline{\epsfbox{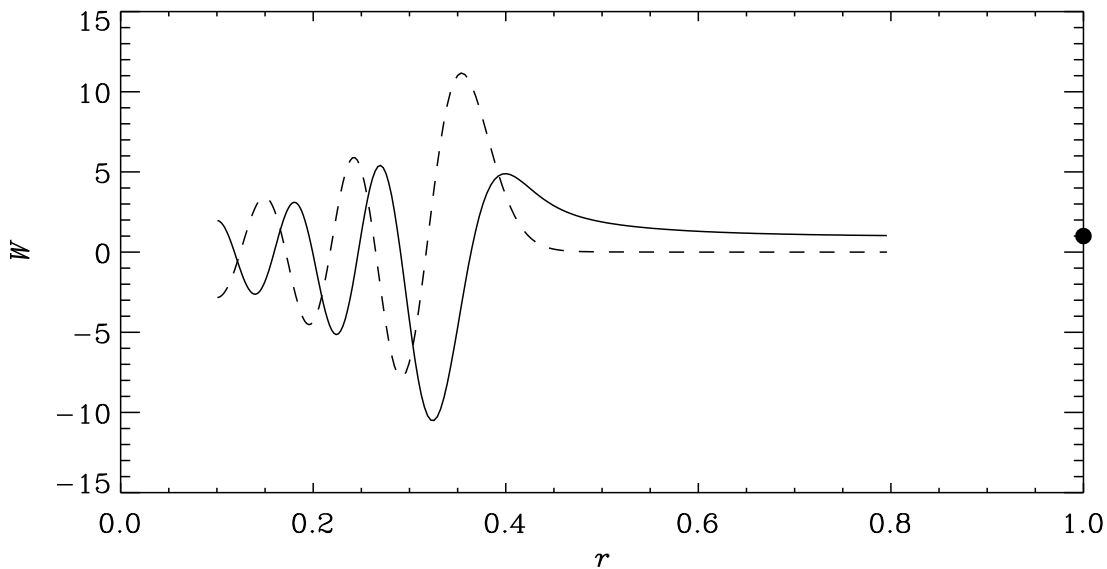}}
  \figcaption[]{Eigenfunction of a
    secular mode for an artificially cold disk with $\alpha=10^{-5}$,
    $\epsilon=10^{-5}$, and $m_{\rm disk}=10^{-6}$.  Real (solid) and
    imaginary (dashed) parts of the complex tilt variable are shown.
    The mode is normalized such that $W_{\rm Jupiter}=1$, while
    $W_{\rm Saturn}\approx-2.47$.  A wave is launched at the location
    of the particle secular resonance ($r=0.381$) due to the
    Jupiter-Saturn interaction. }
\end{figure}

\end{document}